# Emergence of a spin-liquid-like phase in quantum spin-ladder $Ba_2CuTeO_6$ with chemical substitution

Shalini Badola, Devesh Negi, Aprajita Joshi, and Surajit Saha[*]

*Indian Institute of Science Education and Research Bhopal, Bhopal 462066, India.*

*Correspondence: surajit@iiserb.ac.in*

**Abstract:** Stabilization of the quantum spin liquids is vital to realize applications in spintronics and quantum computing. The unique magnetic structure of $Ba_2CuTeO_6$ comprising of coupled spin-ladders with finite inter-ladder coupling brings the system close to the quantum critical point. This opens up possibilities to stabilize unconventional magnetic phases by tailoring the intra- and inter-ladder exchange couplings. Here, we demonstrate a spin-liquid-like phase in $Ba_2CuTeO_6$ using the method of chemical substitution. We choose non-magnetic $La^{3+}$ cation to substitute the $Ba^{2+}$ in $Ba_2CuTeO_6$ and present signature fingerprints such as deprived magnetic transition, non-dispersive AC susceptibility, magnetic field-independent heat capacity, and broad Raman continuum supporting the emergence of a spin-liquid-like phase. We believe that an increased magnetic frustration and spin-fractionalization upon chemical substitution play a crucial role in driving such a state. In addition, temperature and magnetic field-dependent phonon response indicate the presence of magnetostriction (spin-lattice coupling) in La-doped $Ba_2CuTeO_6$, a notable property of spin-liquids.

Spin systems with quantum criticalities hold great potential to exhibit unconventional ground states such as spin-liquids. Quantum spin systems are materials that generally avoid magnetic ordering as a consequence of dominant quantum spin fluctuations. Anderson proposed the notion of enhanced quantum fluctuations in low-spin candidates, particularly the S = ½ and 1 compounds, systems with low-spin connectivity, and frustrated bond networks [1]. The complex interplay between the growing fluctuations (both thermal and spin) and exchange interactions leads to novel ground states in condensed matter systems with strong relevance in spintronics, quantum computing, etc. The novel states have been understood in low spin cuprates as well as highly spin-orbit-coupled iridates with the realization of weak antiferromagnets, spin-liquids, resonance valence bond states along with fractionalized excitations, and unconventional superconductivity [1-4]. In particular, S = ½ systems forming spin-chains, triangular, and ladder-like spin structures are host to unusual fractionalized excitations, leading to quantum spin liquids. However, these states are rare in existence and challenging to stabilize because of a delicate balance between the thermodynamics and disorder. Therefore, low-spin candidates with weak spin connectivity are promising test-beds to stabilize these new exotic phases.

Spin-ladder $Ba_2CuTeO_6$, which lies in the extreme quantum regime of low-spin (S = ½) and weak bond networks, has recently drawn enormous interest of the community [5]. The electronic degeneracy of the $Cu^{2+}$ cation in $CuO_6$ octahedra leads to Jahn-Teller effect which in turn promotes low magnetic dimensionality in the system. In contrast, $TeO_6$ exhibits a strong tendency to build a three-dimensional (*3D*) network. In this conflicting scenario, due to the inter-ladder coupling, the system exists close to the quantum critical point. Previous reports suggest that $Ba_2CuTeO_6$ exhibits a crossover in magnetic dimensionality from the paramagnetic phase to short-ranged quasi-two-dimensional (*2D*) spin-ladder state below ~ 75 K and then to a (*3D*) antiferromagnetic state below ~ 15 K (See Fig.1(a)) [6,7]. The ladder-state, which is the intermediate state, exists in the *bc*-plane with inter-ladder interactions along the *c*-direction (refer Fig.1(b)). The finite leg ($J_l$) and rung ($J_r$) interactions ($J_l /J_r \neq 0$) along with weak inter-ladder coupling extend beyond the two-leg limit and thus stabilizing a *3D* ordered antiferromagnetic network below ~15 K. Theory and experiments have established that even-leg ladder systems may exhibit a spin-liquid state with exponentially decaying spin-spin correlations (SSC) [8, 9]. On the other hand, odd-leg ladders behave akin to spin chain systems with power-law decay SSC [8]. In fact, indications of a possible spin-liquid-like state has been suggested in $Ba_2CuTeO_6$ but only in the spin-ladder state [7]. $Ba_2CuTeO_6$ consists of dominant two-leg spin ladder network with weak inter-ladder coupling. Tailoring the intra- and inter-ladder couplings in $Ba_2CuTeO_6$ can, therefore, introduce uncompensated interactions in spin-lattice, thereby, evolving the quasi-*2D* magnetic character into such unconventional phases.

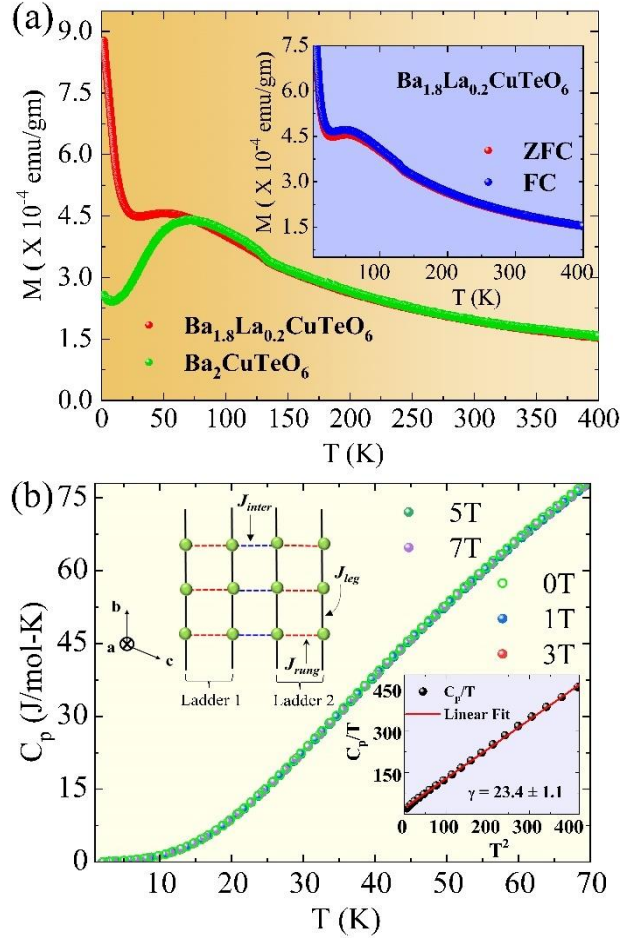

**Figure 1** (Color online): (a) Temperature-dependent Magnetization of the $Ba_2CuTeO_6$ and $Ba_{1.8}La_{0.2}CuTeO_6$. Inset shows that ZFC and FC curves of the $Ba_{1.8}La_{0.2}CuTeO_6$ coincide at all temperatures. (b) Specific heat capacity shown as a function of temperature at variable magnetic fields. Inset displays linear fit to the $C_p/T = \gamma + \beta T^2$.

Chemical substitution is an effective approach to tailor the exchange interactions and establish unconventional but novel phases in condensed matter systems [10,11]. With this motivation, herein, we demonstrate that a quantum spin-liquid-like state can be realized in $Ba_2CuTeO_6$ through the chemical substitution of $Ba^{2+}$ by $La^{3+}$ with a possible charge compensation of $Cu^{2+}$. The signatures of spin-liquid phase are: (i) the suppression of magnetic transition and presence of a non-dispersive magnetic susceptibility in doped system, (ii) magnetic field-independent heat capacity along with the signatures of fermionic density of states, and (iii) the 2-magnon (2M) feature in Raman spectrum of $Ba_2CuTeO_6$ turning into a broad continuum in the La-doped $Ba_2CuTeO_6$ at 5 K. The observed spin-liquid-like behavior in La-doped $Ba_2CuTeO_6$ may originate from fractionalization of spin-singlets and/or an increased magnetic frustration upon chemical substitution. Moreover, we observe anomalies in phonon behavior with temperature and magnetic field in doped $Ba_2CuTeO_6$ that are attributed to spin-phonon (lattice) coupling, which are also notable characteristics of spin-liquids.

Present investigations have been carried out on the pelletized polycrystalline samples of $Ba_2CuTeO_6$ (BCT), $Ba_{1.8}La_{0.2}CuTeO_6$ (BLCT), $Ba_2Cu_{0.8}Zn_{0.2}TeO_6$ (BCZT), and $Ba_2Cu_{0.9}Ni_{0.1}TeO_6$ (BCNT) synthesized using the solid-state reaction method at temperatures 750, 900, and 1050 °C, respectively. The room temperature X-ray diffraction profiles of BCT and BLCT, recorded with PANalytical X-ray Diffractometer were analysed using Rietveld refinement and presented in Fig. S1 [12]. The lattice parameters of both the systems suggest the stabilization of the triclinic phase (Space Group: P-1). Rietveld analysis detects a minor fraction (~ 1.9 %) of the secondary phase of precursor $La_2O_3$ in the doped system BLCT, which seemingly does not impair the magnetism and the other measured properties. Notably, the structural symmetry of BLCT remains unaltered at room temperature despite the substitution of larger divalent $Ba^{2+}$ with smaller trivalent $La^{3+}$ ion. The successful incorporation of $La^{3+}$-ions into the BCT matrix is confirmed using Energy dispersive X-Ray spectroscopy, details of which are provided in the supplemental material [12].

Figure 1 presents the temperature-dependent DC magnetization of the parent and La-doped systems (BCT and BLCT, respectively) recorded using Quantum-Design SQUID-VSM (Superconducting Quantum Interference Device Vibrating Sample Magnetometer) setup. The magnetization of BCT displays a broad feature peaking at $T_S \sim 75$ K, a property common to low-dimensional antiferromagnets, depicting the presence of short-ranged spin-correlations. The broad feature in magnetization intriguingly undergoes a suppression and shifts towards lower temperature upon doping $La^{3+}$ at the $Ba^{2+}$-site (in BLCT) avoiding any discernible spin ordering down to the lowest temperature (~2 K) measured, as shown in Fig.1(a). It is to be noted that $La^{3+}$ is a non-magnetic ion and is not expected to alter the magnetism of BCT so significantly upon doping; yet, smaller size and distinct oxidation of $La^{3+}$ may change the local bond parameters and contribute uncompensated spin interactions to the lattice due to charge disparity at the Ba-site, influencing its magnetic behavior through modification of exchange pathways. A decrease in the magnetic moment may be noted from $(1.8483 \pm 0.0003)\mu_B$ for BCT to $(1.8191 \pm 0.0009)\mu_B$ for BLCT, as detailed in Supplemental Material [12]. Though there is a small change in the magnetic moment, it may be due to the conversion of $Cu^{2+}$ into $Cu^+$ by a small fraction to maintain the charge neutrality of the system [12]. Furthermore, the magnetization of the BLCT exhibits a paramagnetic behavior below $T_P \sim 20$ K which is very weak in the parent system BCT (see Fig.1(a)). Inset in Fig.1(a) shows that the magnetization data recorded under zero field-cooled (ZFC) and field-cooled (FC) protocols for BLCT reveals no sizeable bifurcation in the temperature range under investigation, ruling out the existence of a spin-glass phase [13]. This is further corroborated by the AC magnetic susceptibility ($\chi'$ - real part and $\chi''$ - imaginary part) measurements which revealed a non-dispersive $\chi'$ upon varying the frequency of the applied magnetic field (refer to Fig. S2 in supplemental material [12]). Moreover, $\chi''$ also remained close to zero showing a negligible change under the variable frequency of the applied magnetic field. All of these findings in BLCT are suggestive of a potential spin-liquid behavior and, therefore, call for further measurements to fully understand the distinct characteristics observed in our magnetic studies.

Heat capacity and inelastic light scattering have been two extremely efficient probes to elucidate the nature of spin-spin correlations and their evolution with temperature (T) and magnetic field [14-17]. We performed heat capacity and Raman measurements on the polycrystalline BCT and BLCT with varying temperatures and magnetic fields. The heat capacity, measured using a Physical Property Measurement System (PPMS), reveals that BLCT does not exhibit a sharp λ-like anomaly (see Fig.1(b)), signifying the absence of any long-ranged spin-ordered state down to 2 K. As shown in Fig.1(b), the $C_p$ vs T exhibits a behavior that is independent of applied magnetic field thus confirming the absence of spin-glass phase, as also suggested from our static (DC) and dynamic (AC) magnetic measurements. The heat capacity of BLCT is fitted with $C_p/T = \gamma + \beta T^2$ at low temperatures (2 K < T < 20 K) yielding $\gamma = 23.4 \pm 1.1$ J.mol$^{-1}$K$^{-2}$ and $\beta = 1.05 \pm 0.01$ J.mol$^{-1}$.K$^{-3}$ (see inset in Fig.1(b)). A large magnitude of $\gamma$ (typical values are 1-250 mJ.mol$^{-1}$K$^{-2}$) in the doped BLCT is indicative of contribution from the fermionic density of states (DOS) [13,18]. These states may arise due to quasi-particles introduced through chemical substitution of $Ba^{2+}$ by $La^{3+}$. Notably, $\gamma$ does not change with varying magnetic field, thus ruling out the presence of any paramagnetic impurity [19].

To further understand the characteristic features discussed above, Raman measurements were performed using LabRAM HR Evolution Raman spectrometer attached to attoDRY 1000 He cryostat down to 5 K with an excitation laser source of wavelength 532 nm. Figure 2(a) presents a comparison of the Raman spectra of the parent system BCT and its doped variants at 5 K. The Raman spectrum of BCT demonstrates a broad feature peaking near 190 cm$^{-1}$ which is identified as 2M excitation in an earlier work by Gibbs *et al.* [7] On the contrary, the A-site ($La^{3+}$) doped analogue of BCT (*i.e.*, BLCT) reveals the existence of a broad continuum background (instead of 2M) extended in energy from 0 meV to above ~ 40 meV. To understand the origin of the continuum in BLCT, the Raman spectra of other compositions but with B-site doping (Ni and Zn doping at Cu-site) in BCT (BCNT and BCZT) have been compared with and are shown in Fig.2(a). It is intriguing to note that a 2M feature, similar to that in BCT, remained nearly unaffected and was also observed in BCNT and BCZT at energies ~ 185 cm$^{-1}$ and ~ 160 cm$^{-1}$, respectively. While the substitution of magnetic ($Ni^{2+}$) and non-magnetic ($Zn^{2+}$) ions at the B-site ($Cu^{2+}$) does not destroy the 2M excitation, doping of non-magnetic ion ($La^{3+}$) at the A-site ($Ba^{2+}$) transforms the 2M mode to a broad-continuum. This implies a strong role of A-site substitution on the lattice. A recent report on the inelastic neutron scattering study of BCT revealed that the magnons lie in the *bc*-plane [20] (See Fig.1(b)). Notably, the crystal structure depicts that the Ba-ion (A-site), where chemical substitution is made in this work, exists in between the ladder planes (along the *a*-axis). An anti-site substitution of La-ion at the Te- and Cu-site is highly unlikely due to a large difference in the sizes of the two ions and their oxidation states. Therefore, the origin of a broad continuum instead of 2M mode in BLCT needs proper understanding which is discussed below.

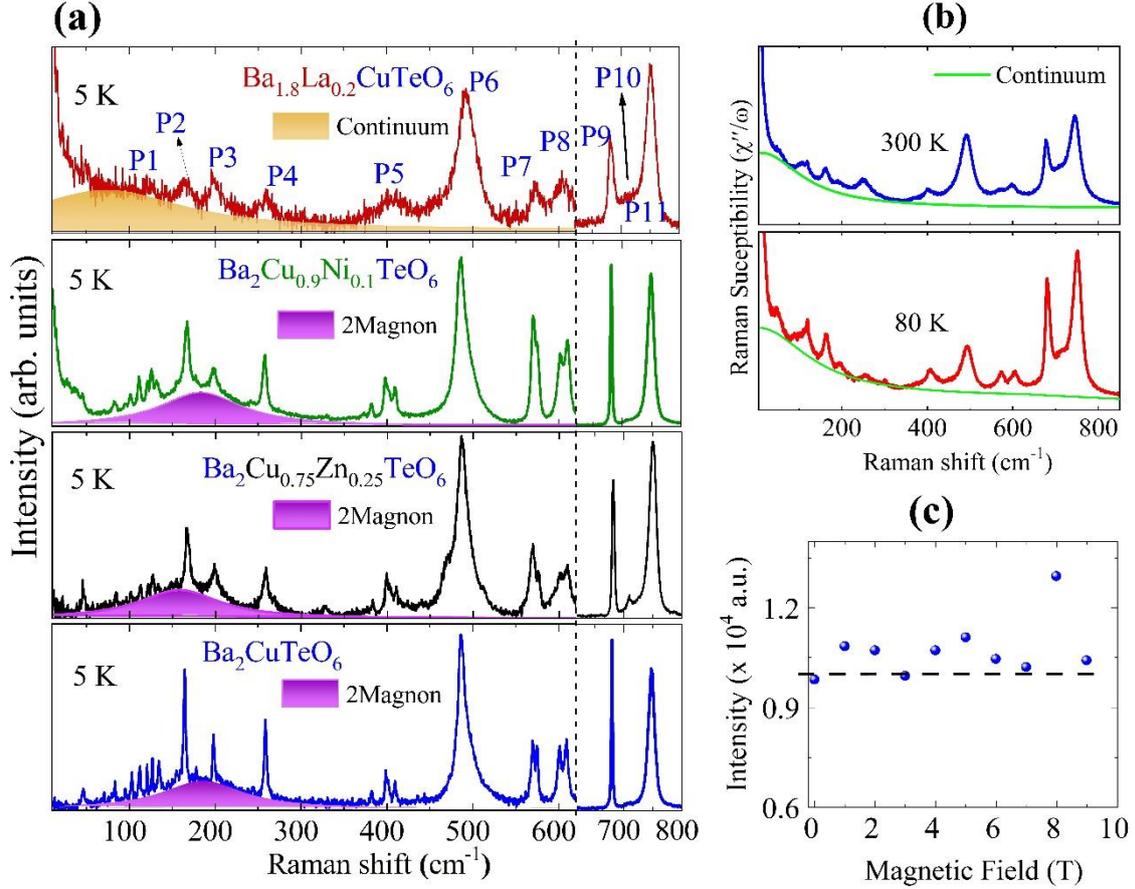

**Figure 2** (Color online): (a) Raman spectrum of parent phase $Ba_2CuTeO_6$ and doped-variants $Ba_{1.8}La_{0.2}CuTeO_6$, $Ba_2Cu_{0.8}Zn_{0.2}TeO_6$, and $Ba_2Cu_{0.9}Ni_{0.1}TeO_6$ showing 2-magnon and broad-continuum at 5 K. Phonons are designated as P1 - P11 at 5 K. (b) Temperature evolution of low-frequency region of $Ba_{1.8}La_{0.2}CuTeO_6$ showing broad feature (shown with green line). (c) Magnetic-field dependent intensity of the continuum in $Ba_{1.8}La_{0.2}CuTeO_6$.

The presence of a broad continuum in the Raman spectrum at low temperatures is intriguing and often associated with the spin, orbital, and/or electronic fluctuations appearing over different temperature and magnetic energy scales. Some of the recent examples of such cases include broad continuum arising due to fractionalized Majorana fermionic excitations in triangular Kitaev magnet α-RuCl$_3$ and β(γ)-Li$_2$IrO$_3$ [2,21], antiferro- and ferro-orbitals in LaMnO$_{3+\delta}$ (0.085≤ δ ≤ 0.125) [22], intervalley fluctuations from single particle excitations in doped Si [23]. We believe that the electronic and orbital origin of the continuum in BLCT is unlikely since its band gap is large (~ 1 eV) and orbital contribution usually occurs at much higher energy scales (typically observed in the range 0.25-2.5 eV), respectively [24]. The absence of a magnetic order (down to 2 K) and the existence of a broad continuum being present down to ~ 0 meV (~ 0 cm$^{-1}$) energy is an indicative of the 2M excitation (S=1) of BCT transforming into a broad spin continuum (S≠1) with La-doping. The analysis of temperature and magnetic field dependence of the continuum reveals intriguing details, as shown in Fig. 2 (b and c). It is observed that the continuum exists even above room temperature. It is worth noting here that such broad Raman continuum has been reported in systems hosting Kitaev spin-liquid behavior, such as Li$_2$IrO$_3$ [2], α-RuCl$_3$ [21], *etc.* and is attributed to arise from fractionalized Majorana excitations (S= ½). Furthermore, our measurements show no effect of the applied magnetic field on the intensity (spectral weight) of this continuum at 5 K (see Fig. 2(c)). Therefore, the observation of the continuum at higher temperatures and its invariance with the magnetic field are indicative of the quantum spin liquid-like correlations in BLCT.

Besides, it must be noted that most of the phonon modes which appear intense and sharp in BCT have merged, become significantly less intense, and broadened in BLCT even at low temperatures (*e.g.*, at 5K) indicating a renormalization of the phonons upon chemical substitution. Among all the phonon modes of BLCT, the magnetic field (*H*) response of the modes at ~ 408 cm$^{-1}$, ~ 490 cm$^{-1}$, and ~ 608 cm$^{-1}$ are notable and shown in Fig. 3. We find that the frequency of these modes shows an increasing trend with an increasing field. Such a field dependence is absent for these phonons in the parent phase BCT (refer to supplemental material [12]). A magnetic field-dependent phonon response is suggestive of the presence of magnetostriction *i.e.,* a coupling between spin and lattice degrees of freedom in BLCT [25].

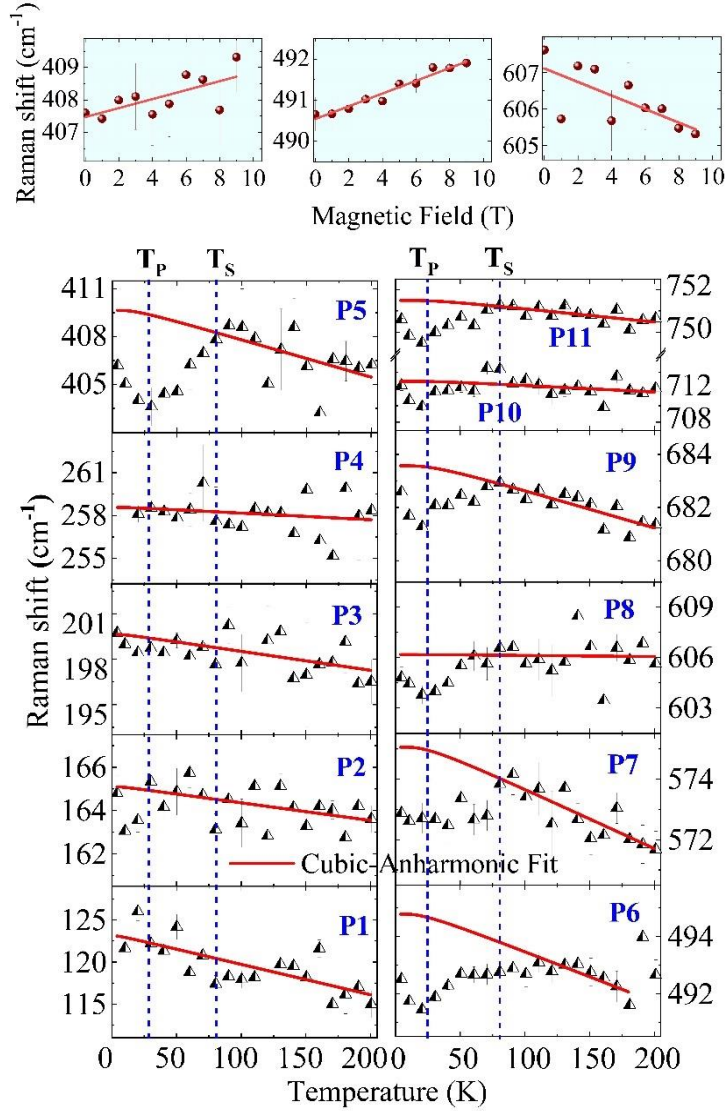

**Figure 3** (Color online): (Top panel) Magnetic field dependence of a few phonons (Lower panel) Phonon frequencies as a function of temperature fitted with cubic-anharmonicity (solid red line). Verticle dashed lines represent the transition temperatures $T_P$ and $T_S$. Error bars are included for phonon frequencies.

On the other hand, the remaining phonon modes show very weak field dependence. A coupled spin-lattice dynamics can also be elucidated from temperature-dependent Raman measurement which reveals that a few phonon modes, especially P5-P11 (see Fig. 3), show an anomalous trend in frequency (ω) upon decreasing the temperature by deviating from conventional cubic-anharmonicity ($\Delta\omega_{anh}(T)$). In the absence of strong magnetic, electronic, and lattice correlations, the phonon behavior is governed by cubic-anharmonic trend which is expressed as [26,27]

$$\Delta\omega^{canh}(T) = \omega_0 + A\left[1 + \frac{2}{e^{\frac{\hbar\omega}{2k_BT}}-1}\right] \quad (1)$$

where $\omega_0$ and $A$ are the fitting parameters, and $\hbar$ and $k_B$ depict the Plancks and Boltzmann constants, respectively. On the other hand, the low-frequency phonons (P1-P4) obey standard anharmonic behavior showing a decreasing frequency with increasing temperature.

A careful observation of Fig. 3 indicates that phonons exhibit anomalies around two temperature ranges upon cooling - firstly below 75 K and then below 20 K. The two temperature ranges in BLCT refer to the short-range ordered magnetic state below $T_S \sim 75$ K and a paramagnetic-like response below $T_P \sim 20$ K, as observed in the magnetization data. The observed anomalies in the phonon response below 75 K and 20 K can be attributed to spin-phonon coupling mediated through modulation of the exchange pathway and interaction with fractionalized excitations, as also observed in Kitaev-$Cu_2IrO_3$ [28].

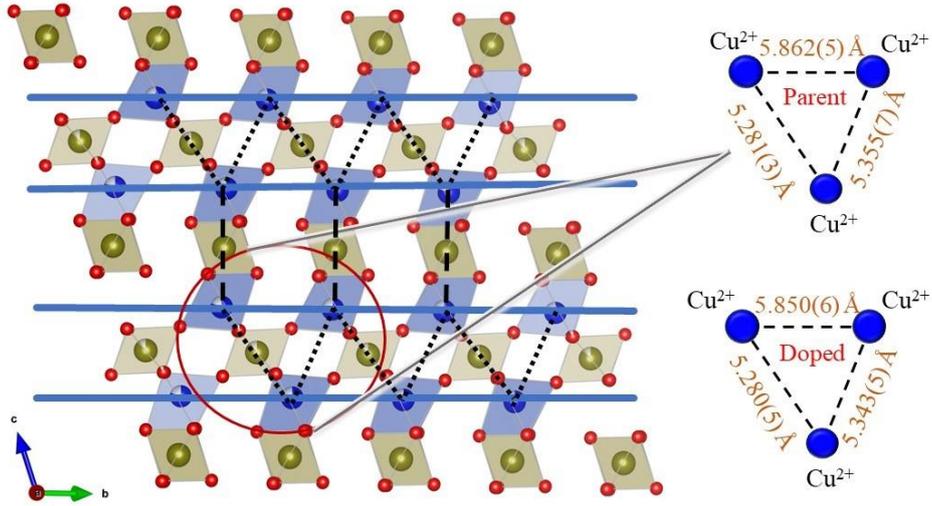

**Figure 4** (Color online): (Left) Schematic depicting spin-ladder configuration in $Ba_{1.8}La_{0.2}CuTeO_6$ where short- and long-dashed lines refer to rung and inter-ladder interactions, respectively. (Right) Cu-Cu separations in the triangular units of spin-ladders of $Ba_2CuTeO_6$ and $Ba_{1.8}La_{0.2}CuTeO_6$.

The signatures of the quantum spin liquid-like state obtained in Magnetic, Specific Heat, and Raman measurements can be understood by considering two scenarios - (1) Based on geometrical factors, especially the in-plane and out-of-plane Cu-Cu separations (extracted using VESTA [29]) and (2) conversion of $Cu^{2+}$ (S= ½) to $Cu^+$ (S=0) in BLCT. In the case of first scenario, we observed that the Cu-Cu separation between the ladder planes (along the *a*-axis) increases upon chemical substitution. On the contrary, the in-plane Cu-Cu separations in

BLCT show decrement as compared to the parent BCT (See Fig. 4). A decreased Cu-Cu separation in the triangular spin arrangement of ladders in BLCT is likely to frustrate the spins more as evidenced in the form of lack of magnetic order. Details of the magnetic frustration in BCT and BLCT are provided in Fig.4 and Supplemental Material [12]. In the other scenario, an oxidation state conversion from $Cu^{2+}$ (S= ½) to $Cu^+$ (S=0) in BLCT due to doping by a trivalent $La^{3+}$ at the A-site. The inverse susceptibility data suggest comparable magnetic moments for the two systems (BCT: 1.84$\mu_B$ and BLCT: 1.82$\mu_B$) (see Supplemental Material [12]) implying the conversion of a small fraction of the Cu oxidation state. Even though small, unsatisfied Cu interactions will produce fractionalized quasi-particles (S = 1/2) in the lattice which are formed out of broken spin-pair interactions between the spin-ladders and may lead to the observed continuum in Raman spectra. Thus, we believe that an increased magnetic frustration and fractionalized excitations contribute to and explain the spin-liquid-like behavior in BLCT.

In conclusion, a quantum spin-liquid-like state has been realized in chemically modified $Ba_2CuTeO_6$. Several signatures such as lack of magnetic order and spin-glass phase, fermionic density of states in specific heat, and broad continuum in Raman measurement at low and high temperatures were identified to support the existence of the spin-liquid-like state in $Ba_{1.8}La_{0.2}CuTeO_6$. Two approaches are described to explain the origin of liquid-like correlations - one with an increased magnetic frustration due to lowered Cu-Cu separation in the lattice and the other with an oxidation state conversion of $Cu^{2+}$ into diamagnetic $Cu^+$ ions upon $La^{3+}$-doping leading to fractionalized quasi-particles in the system. However, *μSR* investigations at low temperatures can be useful to ascertain this suggestion which is beyond the scope of the present work and can be left for future investigations.

## Acknowledgements

SS acknowledges Science and Engineering Research Board (SERB) for funding through ECR/2016/001376 and CRG/2019/002668. Funding from DST-FIST (Project No. SR/FST/PSI-195/2014(C)) and Nano-mission (Project No. SR/NM/NS-84/2016(C)) are also acknowledged. D. N. acknowledges the fellowship (09/1020(0139)/2018-EMR-I) support from CSIR. A. J. acknowledges the CSIR fellowship (09/1020(0179)/2019-EMR-I). Authors acknowledge Central Instrumentation Facility at IISER Bhopal for temperature-dependent XRD and SQUID-VSM facilities.

## References:

[1] P. W. Anderson, The resonating valence bond state in $La_2CuO_4$ and superconductivity, Science **235**, 1196(1987).

[2] A. Glamazda, P. Lemmens, S.-H. Do, Y. Choi, and K.-Y. Choi, Raman spectroscopic signature of fractionalized excitations in the harmonic-honeycomb iridates β-and γ-$Li_2IrO_3$, Nature Communications **7**, 12286 (2016).

[3] Y. Singh, S. Manni, J. Reuther, T. Berlijn, R. Thomale, W. Ku, S. Trebst, and P. Gegenwart, Relevance of the heisenberg-kitaev model for the honeycomb lattice iri-dates $A_2IrO_3$, Physical Review Letters **108**, 127203 (2012).

[4] M.-K. Wu, J. R. Ashburn, C. Torng, P.-H. Hor, R. L. Meng, L. Gao, Z. J. Huang, Y. Wang, and C. W. Chu, Superconductivity at 93 k in a new mixed-phase Y-Ba-Cu-O compound system at ambient pressure, Physical Review Letters **58**, 908 (1987).

[5] A. Gibbs, A. Yamamoto, A. Yaresko, K. Knight, H. Yasuoka, M. Majumder, M. Baenitz, P. Saines, J. Hester, D. Hashizume, et al., S = 1/2 quantum critical spin ladders produced by orbital ordering in $Ba_2CuTeO_6$, Physical Review B **95**, 104428 (2017).

[6] G. N. Rao, R. Sankar, A. Singh, I. P. Muthuselvam, W. Chen, V. N. Singh, G.-Y. Guo, and F. Chou, Tellurium-bridged two-leg spin ladder in $Ba_2CuTeO_6$, Physical Review B **93**, 104401 (2016).

[7] A. Glamazda, Y. Choi, S.-H. Do, S. Lee, P. Lemmens, A. Ponomaryov, S. Zvyagin, J. Wosnitza, D. P. Sari, I. Watanabe, et al., Quantum criticality in the coupled two-leg spin ladder $Ba_2CuTeO_6$, Physical Review B **95**, 184430 (2017).

[8] T. Saha-Dasgupta, The fascinating world of low-dimensional quantum spin systems: Ab initio modeling, Molecules **26**, 1522 (2021).

[9] B. Normand and T. Rice, Electronic and magnetic structure of $LaCuO_{2.5}$, Physical Review B **54**, 7180 (1996).

[10] Z. Pan, J. Chen, X. Jiang, L. Hu, R. Yu, H. Yamamoto, T. Ogata, Y. Hattori, F. Guo, X. Fan, et al., Colossal volume contraction in strong polar perovskites of $Pb(Ti,V)O_3$, Journal of the American Chemical Society **139**, 14865 (2017).

[11] J.-M. Tarascon, P. Barboux, P. Miceli, L. Greene, G. Hull, M. Eibschutz, and S. Sunshine, Structural and physical properties of the metal (m) substituted $YBa_2Cu_{3-x}M_xO_{7-y}$ perovskite, Physical Review B **37**, 7458 (1988).

[12] S. Badola, D. Negi, A. Joshi, and S. Saha, This supplemental material includes details of refined X-ray diffraction data of $Ba_2CuTeO_6$ and $Ba_{1.8}La_{0.2}CuTeO_6$, energy dispersive x-ray spectroscopy, ac magnetic susceptibility, estimation of magnetic moments in $Ba_{1.8}La_{0.2}CuTeO_6$, and magnetic field dependent Raman data of $Ba_2CuTeO_6$.

[13] O. Mustonen, S. Vasala, E. Sadrollahi, K. Schmidt, C. Baines, H. Walker, I. Terasaki, F. Litterst, E. Baggio-Saitovitch, and M. Karppinen, Spin-liquid-like state in a spin-½ square

lattice antiferromagnet perovskite induced by $d^{10}$–$d^0$ cation mixing, Nature communications **9**, 1085 (2018).

[14] S. Bramwell, M. Harris, B. Den Hertog, M. Gingras, J. Gardner, D. McMorrow, A. Wildes, A. Cornelius, J. Champion, R. Melko, *et al.*, Spin correlations in $Ho_2Ti_2O_7$: a dipolar spin ice system, Physical Review Letters **87**, 047205 (2001).

[15] J. Laverdière, S. Jandl, A. Mukhin, V. Y. Ivanov, V. Ivanov, and M. Iliev, Spin-phonon coupling in orthorhombic $RMnO_3$ (R = Pr, Nd, Sm, Eu, Gd, Tb, Dy, Ho, Y): A raman study, Physical Review B **73**, 214301 (2006).

[16] S. Li, Z. Ye, X. Luo, G. Ye, H. H. Kim, B. Yang, S. Tian, C. Li, H. Lei, A. W. Tsen, et al., Magnetic-field-induced quantum phase transitions in a van der waals magnet, Physical Review X **10**, 011075 (2020).

[17] J. Xing, L. D. Sanjeewa, J. Kim, G. Stewart, A. Podlesnyak, and A. S. Sefat, Field-induced magnetic transition and spin fluctuations in the quantum spin-liquid candidate $CsYbSe_2$, Physical Review B **100**, 220407 (2019).

[18] L. Balents, Spin liquids in frustrated magnets, Nature **464**, 199 (2010).

[19] S. Yamashita, T. Yamamoto, Y. Nakazawa, M. Tamura, and R. Kato, Gapless spin liquid of an organic triangular compound evidenced by thermodynamic measurements, Nature communications **2**, 275 (2011).

[20] D. Macdougal, A. S. Gibbs, T. Ying, S. Wessel, H. C. Walker, D. Voneshen, F. Mila, H. Takagi, and R. Coldea, Spin dynamics of coupled spin ladders near quantum criticality in $Ba_2CuTeO_6$, Physical Review B **98**, 174410 (2018).

[21] L. J. Sandilands, Y. Tian, K. W. Plumb, Y.-J. Kim, and K. S. Burch, Scattering continuum and possible fractionalized excitations in α-$RuCl_3$, Physical Review Letters **114**, 147201 (2015).

[22] K.-Y. Choi, Y. G. Pashkevich, V. Gnezdilov, G. Güntherodt, A. Yeremenko, D. Nabok, V. Kamenev, S. Barilo, S. Shiryaev, A. Soldatov, et al., Orbital fluctuating state in ferromagnetic insulating $LaMnO_{3+\delta}$ ($0.085 \leq \delta \leq 0.125$) studied using raman spectroscopy, Physical Review B **74**, 064406 (2006).

[23] K. Jain, S. Lai, and M. V. Klein, Electronic raman scattering and the metal-insulator transition in doped silicon, Physical review B **13**, 5448 (1976).

[24] R. Rückamp, E. Benckiser, M. Haverkort, H. Roth, T. Lorenz, A. Freimuth, L. Jongen, A. Möller, G. Meyer, P. Reutler, *et al.*, Optical study of orbital excitations in transition-metal oxides, New Journal of Physics **7**, 144 (2005).

[25] B. Poojitha, A. Rathore, A. Kumar, and S. Saha, Signatures of magnetostriction and spin-phonon coupling in magnetoelectric hexagonal 15R-$BaMnO_3$, Physical Review B **102**, 134436 (2020).

[26] M. Balkanski, R. Wallis, and E. Haro, Anharmonic effects in light scattering due to optical phonons in silicon, Physical Review B **28**, 1928 (1983).

[27] S. Saha, D. S. Muthu, S. Singh, B. Dkhil, R. Suryanarayanan, G. Dhalenne, H. Poswal, S. Karmakar, S. M. Sharma, A. Revcolevschi, et al., Low-temperature and high-pressure raman and x-ray studies of pyrochlore $Tb_2Ti_2O_7$: phonon anomalies and possible phase transition, Physical Review B **79**, 134112 (2009).

[28] S. Pal, A. Seth, P. Sakrikar, A. Ali, S. Bhattacharjee, D. Muthu, Y. Singh, and A. Sood, Probing signatures of fractionalization in the candidate quantum spin liquid $Cu_2IrO_3$ via anomalous raman scattering, Physical Review B **104**, 184420 (2021).

[29] K. Momma and F. Izumi, Vesta 3 for three-dimensional visualization of crystal, volumetric and morphology data, Journal of applied crystallography **44**, 1272 (2011).

# Supplemental Material

# Emergence of a spin-liquid-like phase in quantum spin-ladder $Ba_2CuTeO_6$ with chemical substitution

Shalini Badola, Devesh Negi, Aprajita Joshi, and Surajit Saha[*]

*Indian Institute of Science Education and Research Bhopal, Bhopal 462066, India.*

*\*Correspondence: surajit@iiserb.ac.in*

This supplemental material includes details of refined X-ray diffraction data of $Ba_2CuTeO_6$ and $Ba_{1.8}La_{0.2}CuTeO_6$, energy dispersive X-ray spectroscopy, AC magnetic susceptibility, estimation of magnetic moments in $Ba_{1.8}La_{0.2}CuTeO_6$, and magnetic field dependent Raman data of $Ba_2CuTeO_6$.

# SI. X-Ray Diffraction and Energy Dispersive X-Ray spectroscopy

Room temperature X-Ray diffraction profiles of $Ba_2CuTeO_6$ and $Ba_{1.8}La_{0.2}CuTeO_6$ were refined using Highscore Plus. The refinement suggests a triclinic crystal structure for both the polycrystalline samples. Lattice constants of the triclinic phase were extracted and shown in Fig. S1 for both the samples. The peak corresponding to the second phase ($La_2O_3$) is encircled in purple color. Further, it can be observed that the lattice constants $b$ and $c$ decrease upon chemical substitution of Ba by La. On the other hand, the parameter $a$ shows an expansion upon chemical substitution. An estimation of atomic percentages of elements is performed using Energy Dispersive X-ray Analysis (EDAX), and are listed in Table SI below. The measurement was taken at several spots to confirm the sample stoichiometry.

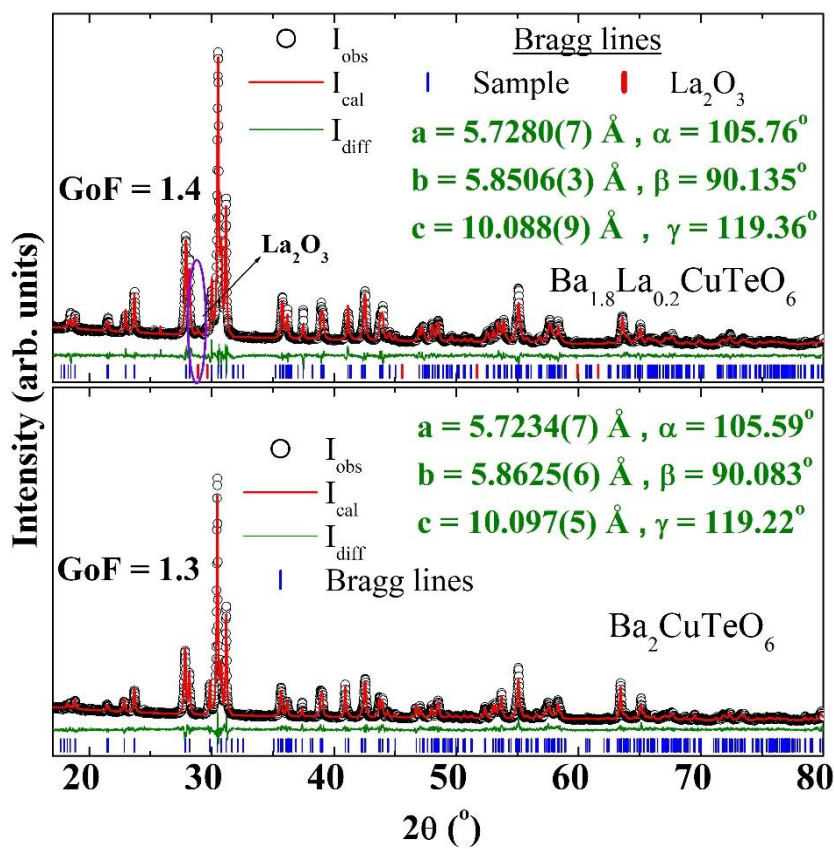

**Figure S1** (Color online): Rietveld refined room-temperature X-ray diffraction profiles of $Ba_2CuTeO_6$ and $Ba_{1.8}La_{0.2}CuTeO_6$ showing the lattice parameters. The peak encircled in purple in the top panel is for the secondary phase of $La_2O_3$.

Table SI. Compositional percentage of elements in $Ba_{1.8}La_{0.2}CuTeO_6$.

| Spot | Atomic percentage of Elements | | | | |
|---|---|---|---|---|---|
| | Ba | La | Cu | Te | O |
| 1 | 13.61 | 1.12 | 7.43 | 8.02 | 69.82 |
| 2 | 15.70 | 1.89 | 9.01 | 9.22 | 64.17 |
| 3 | 14.56 | 2.23 | 7.44 | 7.89 | 67.88 |
| 4 | 14.74 | 2.24 | 7.55 | 7.61 | 67.86 |
| 5 | 15.93 | 2.53 | 7.80 | 8.09 | 65.64 |
| 6 | 15.61 | 2.03 | 9.39 | 8.84 | 64.13 |
| 7 | 14.96 | 2.24 | 8.41 | 7.96 | 66.44 |
| Average | 15.02 | 2.04 | 8.14 | 8.23 | 66.56 |

**SII. AC Magnetic Susceptibility**

AC magnetic susceptibility measurement of $Ba_{1.8}La_{0.2}CuTeO_6$ was taken at multiple frequencies in the temperature range of 2-200 K, as shown in Fig. S2. It is observed that the real part ($\chi'$) of the magnetic susceptibility remains invariant to varying frequencies showing a broad peak at $\sim 75$ K, in agreement with the DC susceptibility. The imaginary part ($\chi''$) also shows no change with varying frequencies. This implies the absence of spin-glass phase in $Ba_{1.8}La_{0.2}CuTeO_6$.

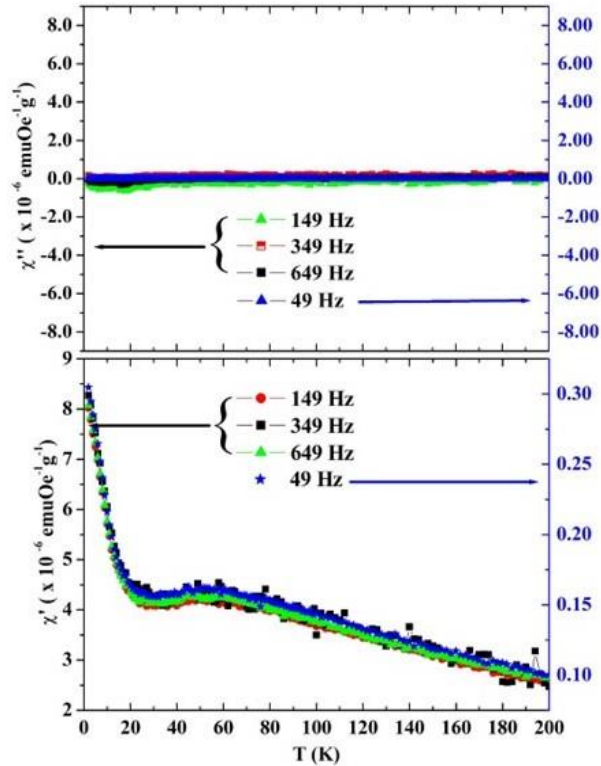

**Figure S2** (Color online): AC magnetic susceptibility of $Ba_{1.8}La_{0.2}CuTeO_6$ showing the variation of the real and imaginary part as a function temperature at different frequencies.

## SIII. Analysis of Inverse Magnetic Susceptibility and Magnetic frustration

Inverse magnetic susceptibility ($\chi^{-1}$) as a function of temperature (T) is plotted for the parent $Ba_2CuTeO_6$ and doped $Ba_{1.8}La_{0.2}CuTeO_6$ compounds. The inverse susceptibility was fitted using the Curie-Weiss law ($\chi = \frac{C}{T-\theta_{CW}}$) where C refers to curie-constant and $\theta_{CW}$ is curie-weiss temperature) for both the systems and the corresponding Curie-Weiss temperatures were obtained as shown in Fig. S3. An antiferromagnetic transition for $Ba_2CuTeO_6$ occurs at $T_N$ = 15 K, as reported earlier [1-3]. On the other hand, the estimated $\theta_{CW}$ for $Ba_2CuTeO_6$ is -87 K. This yields a frustration parameter (f = $\frac{|\theta_{CW}|}{T_N}$) of 6. The magnetic susceptibility recorded for the doped $Ba_{1.8}La_{0.2}CuTeO_6$ does not show any magnetic ordering suggesting that the system exhibits a strong magnetic frustration (*i.e.,* larger frustration parameter compared to $Ba_2CuTeO_6$) as also depicted from the analysis of Cu-Cu separation in the main text. Further, an estimation of the effective magnetic moments ($\mu_{eff}$) for the two systems were extracted from the fitting parameter (C) obtained from the Curie-Weiss fit using the relation $\mu_{eff} = \sqrt{8\,C}$. The effective magnetic moments are estimated to be 1.84$\mu_B$ and 1.82$\mu_B$ for $Ba_2CuTeO_6$ and $Ba_{1.8}La_{0.2}CuTeO_6$, respectively, implying a small fraction of the change (~ 1%) in the Cu oxidation state.

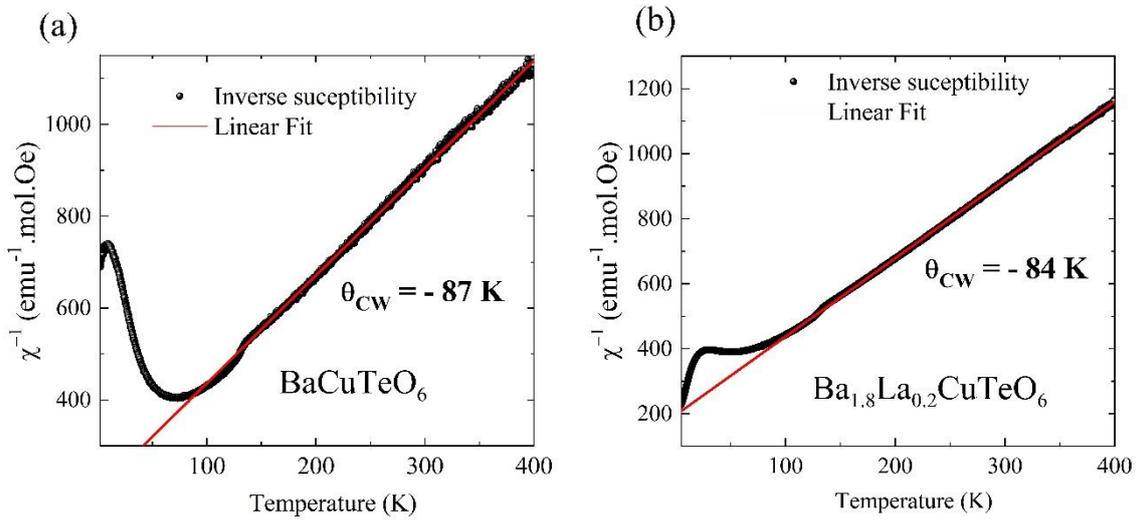

**Figure S3** (Color online): Inverse magnetic susceptibility of (a) $Ba_2CuTeO_6$ and (b) $Ba_{1.8}La_{0.2}CuTeO_6$ showing Curie-Weiss fit.

As we know that a secondary phase also exists in the doped sample. However, it is to be noted that the secondary phase is not expected to influence the magnetic behavior of the primary phase as $La_2O_3$ exhibits negative magnetization in the temperature range of 2-250 K, as shown in the Fig. S4.

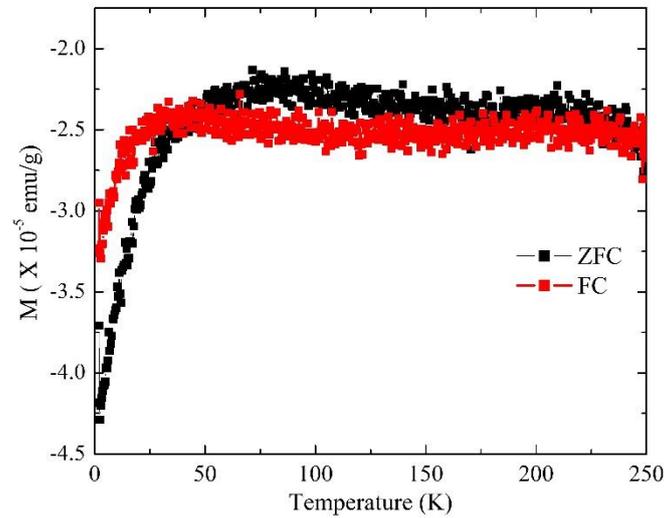

**Figure S4** (Color online): Temperature-dependent magnetization of $La_2O_3$.

### SIV. MAGNETIC FIELD DEPENDENT RAMAN ANALYSIS

Raman measurements were performed in the magnetic field range of 0-9T for the parent $Ba_2CuTeO_6$ as well as doped $Ba_{1.8}La_{0.2}CuTeO_6$ systems. It is observed that a few phonon modes (P5, P6, and P8) respond to the varying field in the doped system as shown in the main text. On the contrary, the same phonon modes do not demonstrate any noticeable variation with changing magnetic fields, as shown in Fig. S5. Thus, a response of the phonon frequencies to a varying magnetic field is observed only in the spin-liquid system depicting a spin-lattice correlation.

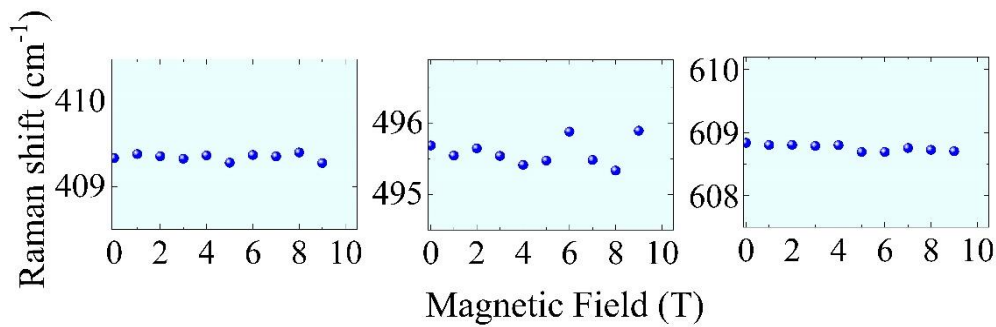

**Figure S5** (Color online): Phonon frequencies as a function of magnetic field in $Ba_2CuTeO_6$.

**References:**

[1] A. Glamazda, P. Lemmens, S.-H. Do, Y. Choi, and K.-Y. Choi, Raman spectroscopic signature of fractionalized excitations in the harmonic-honeycomb iridates β-and γ-$Li_2IrO_3$, Nature Communications **7**, 12286 (2016).

[2] A. Glamazda, Y. Choi, S.-H. Do, S. Lee, P. Lemmens, A. Ponomaryov, S. Zvyagin, J. Wosnitza, D. P. Sari, I. Watanabe, *et al.*, Quantum criticality in the coupled two-leg spin ladder $Ba_2CuTeO_6$, Physical Review B **95**, 184430 (2017).

[3] G. N. Rao, R. Sankar, A. Singh, I. P. Muthuselvam, W. Chen, V. N. Singh, G.-Y. Guo, and F. Chou, Tellurium-bridged two-leg spin ladder in $Ba_2CuTeO_6$, Physical Review B **93**, 104401 (2016).